\documentclass[conference]{IEEEtran}
\IEEEoverridecommandlockouts
\usepackage{cite}
\usepackage{amsmath,amssymb,amsfonts}
\usepackage{graphicx}
\usepackage{textcomp}
\usepackage{xcolor}
\usepackage{cite}
\usepackage{listings}
\usepackage{tabularx}
\usepackage{booktabs}
\usepackage{xcolor, xparse}
\usepackage{algorithm}
\usepackage[noend]{algpseudocode}
\newcommand{\ra}[1]{\renewcommand{\arraystretch}{#1}}

\def\BibTeX{{\rm B\kern-.05em{\sc i\kern-.025em b}\kern-.08em
    T\kern-.1667em\lower.7ex\hbox{E}\kern-.125emX}}

\usepackage{enumitem}
\usepackage{noindentafter}
\NoIndentAfterEnv{itemize}
\NoIndentAfterEnv{description}
\NoIndentAfterEnv{enumerate}
\setlist{
	leftmargin=9pt,
	parsep=2pt plus 1pt,
	listparindent=0pt,%10pt\parindentamount,
}
\setlist[description]{
	font=\normalfont\itshape}
\colorlet{punct}{red!60!black}
\definecolor{background}{HTML}{EEEEEE}
\definecolor{delim}{RGB}{20,105,176}
\colorlet{numb}{magenta!60!black}
 
\lstdefinelanguage{json}{
    numbersep=5pt,
    frame=lines,
    backgroundcolor=\color{background},
    literate=
     *{0}{{{\color{numb}0}}}{1}
      {1}{{{\color{numb}1}}}{1}
      {2}{{{\color{numb}2}}}{1}
      {3}{{{\color{numb}3}}}{1}
      {4}{{{\color{numb}4}}}{1}
      {5}{{{\color{numb}5}}}{1}
      {6}{{{\color{numb}6}}}{1}
      {7}{{{\color{numb}7}}}{1}
      {8}{{{\color{numb}8}}}{1}
      {9}{{{\color{numb}9}}}{1}
      {:}{{{\color{punct}{:}}}}{1}
      {,}{{{\color{punct}{,}}}}{1}
      {\{}{{{\color{delim}{\{}}}}{1}
      {\}}{{{\color{delim}{\}}}}}{1}
      {[}{{{\color{delim}{[}}}}{1}
      {]}{{{\color{delim}{]}}}}{1},
}
\begin{document}

\title{Scalable Call Graph Constructor for Maven\\}

\author{\IEEEauthorblockN{1\textsuperscript{st} Mehdi Keshani \\ \textit{\small{Technical University of Delft, m.keshani@tudelft.nl}}}
}

\maketitle

\begin{abstract}
As a rich source of data, 
Call Graphs are used for various applications 
including security vulnerability detection.
Despite multiple studies showing that Call Graphs 
can drastically improve the accuracy of analysis, 
existing ecosystem-scale tools like Dependabot 
do not use Call Graphs and work at the package-level. 
Using Call Graphs in ecosystem use cases is not practical 
because of the scalability problems that Call Graph generators have. 
Call Graph generation is usually considered to be a ``full program analysis'' 
resulting in large Call Graphs and expensive computation. 
To make an analysis applicable to ecosystem scale, 
this pragmatic approach does not work, 
because the number of possible combinations of 
how a particular artifact can be combined in a full program explodes. 
Therefore, it is necessary to make the analysis incremental.
There are existing studies on different types of incremental program analysis. 
However, none of them focuses on Call Graph generation for an entire ecosystem.
In this paper, we propose an incremental implementation of the CHA algorithm 
that can generate Call Graphs on-demand, 
by stitching together partial Call Graphs 
that have been extracted for libraries before.
Our preliminary evaluation results show 
that the proposed approach scales well and 
outperforms the most scalable existing framework called OPAL.

\end{abstract}


\begin{table*}[h!]
\ra{0.9}
\scriptsize
\caption{Time of CG generation different phases.}
\centering
\label{tab:deps}
\begin{tabular}{@{}rlrrrrrr@{}}
\toprule
 & \textbf{Maven Coordinate}                         & \textbf{\#Deps}          & \textbf{OPAL} & \textbf{CG Pool(1)} & \textbf{Stitching(2)} & \textbf{UCH(3)} & \textbf{1+2+3}\\ \midrule
1 & \scriptsize{com.google.code.maven-play-plugin.org.playframework:play:1.3.2} & 61                       & 2:39min                             & 0:54min                                   & 0:55min                                      & 330ms                                  & 1:50min                     \\ 
... &\multicolumn{7}{r}{...}\\ 
9 & \scriptsize{org.apache.solr:solr-map-reduce:5.4.1}                          & 121                       & 5:03min                             & 1:15min                                    & 3:33min                                     & 459ms                                  & 4:49min                     \\ 
10 & \scriptsize{org.digidoc4j:digidoc4j:1.0.8.beta.2}                          & 49                        & 0:34min                              & 0:21min                                    & 0:13min                                      & 106ms                                  & 0:35min                      \\ \midrule[1.2pt]
 &\scriptsize{First round of generation excluding redundant deps}                                                 & 605  & \textbf{18:46min}                            & 4:33min                                   & 8:32min                                     & 0:02min                                 & \textbf{13:08min}                    \\ 
 & \scriptsize{+Second round of generation} & 605  & \textbf{18:46min} & 0:00min                                   & 8:32min                                     & 0:02min                                 & \textbf{8:34min}                     \\ \bottomrule
\end{tabular}
\end{table*}

\section{Introduction}

In modern Software Engineering, 
the choice of the programming language is as important as the surrounding ecosystem. 
Many tools and reusable components exist 
that make developers more productive.
Software ecosystems ease the management of third-party
libraries. They pull in the dependencies
on demand when necessary.
Maven is a popular ecosystem that hosts more than six million software artifacts.
However, importing dependencies into a project 
also introduces risks like security vulnerabilities of the dependency.
On the other hand, 
fine-grained analysis can have positive impacts on 
reliability of software reuse in ecosystems by improving the accuracy of analyses such as vulnerability detection~\cite{boldi2020fine, hejderup2018software}.
Such fine-grained analysis needs to be performed on Call Graphs (CGs). 
The common approach for constructing a CG is to provide a complete application, 
including all of its dependencies for the CG algorithm. 
However, this approach is not practical for an entire ecosystem. 
It is not scalable due to redundant computations.
The main challenges that 
cause redundancy in ecosystem CG generation are as follows:
\label{popLibs}
(1) Existing Java CG generators 
generate a CG for a given ClassPath (CP). 
Suppose we want to generate CGs for an ecosystem, 
we have to provide the CP of all packages
that exist in the ecosystem.
These CPs also include the libraries that each package uses.
On the other hand, \textit{"a majority of packages 
depends on a small minority of other packages"}\cite{decan2019empirical}. 
Moreover, different versions of a package,
especially if they are minor releases apart,
share a lot of similar dependencies.
Therefore, an ecosystem like Maven, 
results in constructing the same CG again and again 
for each popular library. 
(2) Version range dependency specification on Maven
can cause non-deterministic dependency sets. 
If there is a version range dependency specification in a package 
the result of the dependency resolution may be different 
based on the time of the resolution~\cite{hejderup2018pr}. 
Moreover, various resolution rules of companies or different package managers also make resolution results diverse.
Transitive dependencies can also make dependency sets non-deterministic. 
If direct dependencies of an application have one transitive dependency in common 
and they use different versions of it, 
there will be a version conflict.
The resolver solves this based on the resolution policies. 
Maven chooses the closest version to the root package. 
In any case, if the resolved dependency set slightly changes 
the resulting CG will be different for the same package, 
hence a new CG generation needs to be triggered from scratch 
unless we pre-compute the common parts.
\label{ecosystemCases}
(3) On-demand analysis on top of CGs 
such as fine-grained vulnerability analysis
is time-consuming and expensive.
Such analyses are useful for developers or library maintainers.
An analysis provider needs to load binary files into memory
and construct CGs for them 
which is overly expensive.
Additionally, duplicate computations lower the performance of query responses.
For example, 
if two clients query the server 
of an analysis provider at the same time 
and both of them are using 
\textit{log4j:log4j:jar:1.2.17} library, 
the server has to construct CG for this library
twice at the same time.

In this paper, 
we propose an incremental CG construction approach that 
makes ecosystem-scale CG generation achievable.
Although there are a few studies 
\cite{souter2001incremental, toman2017taming, arzt2016stubdroid} 
on incremental program analysis,
to the best of our knowledge, 
none of them constructs CGs in the scale of the entire Java ecosystem.

We exploit the Maven central ecosystem in this study.
Our approach has three main steps.
First, we construct and store partial CGs for packages
without their dependencies. 
And then, we stitch them whenever it is needed.
Although in this paper we focus on the Maven and CG generation for Java, 
the idea of pre-computation per package can be used for other ecosystems.
We use the OPAL CG generator to generate the partial CGs. 
We also compare our results with this framework as a baseline.
Our evaluation results show that 
the proposed approach
can highly affect the scalability of CG generation 
and outperform the most scalable 
existing framework, OPAL~\cite{reif2019judge}.
The main contribution of this work is a novel CG generation technique that;
(1) makes CG generation possible for Maven ecosystems, 
(2) improves the scalability of existing approaches, 
(3) removes redundant computation from CG generation, and finally
(4) enables efficient and on-demand CG-based analysis.

\section{Related Work}
\label{related}

There are several studies on the scalability of different analyses.
Tip et al. \cite{tip2000scalable} proposed various algorithms 
to improve the scalability of CG generation.
However, in their study, they focus on 
large programs, not an entire ecosystem.
Alexandru et al. \cite{alexandru2019redundancy} 
took advantage of the same concept that we do,
which is avoiding multiple analysis of redundant parts.
However, CG generation is not the focus of their study. 

There also exist several studies on incremental static analysis.
Souter et al.\cite{souter2001incremental} 
made the CPA algorithm incremental. 
Their proposed approach updates a CG with the changed parts of new versions.
However, the scale of their work is multiple releases, not an ecosystem.  
Arzt et al. \cite{arzt2016stubdroid} uses summary pre-computation 
to improve the scalability of data flow analysis on android applications.
Although the pre-computation is very relevant to our approach, 
they do not use it for CG generation.  
To the best of our knowledge, 
no existing study uses pre-computation of packages to generate ecosystem-scale CGs.

\section{Methodology}
As opposed to existing approaches, 
we propose to remove dependency resolution as the pre-step of CG construction.
We untangle the resolution process and CG construction by using the dependency set 
as a parameter of CG construction.
Therefore, we generate and store a partial CG for 
each package only once and use it many times in the future.
Considering that CG construction 
is a heavy computation and resulting CGs 
are mostly heavy objects\footnote{Objects that occupy a lot of memory in the program}, 
by removing duplications from the process 
we save a lot of time and storage.

In the proposed approach,
we first download binaries of packages from Maven Central.
Then, we generate CGs for them using an existing CG generator.
Next, we parse the output of the CG generator and 
extract concise yet sufficient information for further steps.
This information includes class hierarchy and 
CG information of the package and will be stored
in a storage called \textit{CG Pool}.
In the CG Pool, each CG is indexed by its package reference
which in the case of Maven is a \textit{Maven Coordinate}.
A Maven Coordinate is composed of \textit{groupId:artifactId:version},
that uniquely identifies a package within the whole ecosystem.
This CG pool can be updated, 
whenever a new release is out on Maven.
After we create the CG pool,
any custom dependency set
can be used to generate a full CG.
Whenever we have a set of packages 
as a result of a resolution,
we fetch the computed CGs from the CG pool and
stitch them together using the algorithm that we have implemented.

Once we have a resolution set, we combine the CGs 
that we have previously fetched from the CG pool and create a Universal Class Hierarchy (UCH).
Then the Stitching algorithm walks through the edges of CGs,
and based on the type of the invocation\footnote{There are five types of invocations in the JVM bytecode e.g. invokestatic} 
decides how to resolve new edges or correct the existing edges.
That is, edges that we have in the CGs of the CG pool,
are not complete due to lack of information in partial CG construction.
Hence, stitching tries to complete these edges 
by adding new edges to libraries or replacing the existing edges that are incorrect.

\section{Results and Plans}

We implemented a prototype of the proposed solution 
and we observed the expected improvements in the scalability.
We compared the execution time of OPAL 
and the proposed solution on ten dependency sets.
As shown in the Table~\ref{tab:deps}, 
OPAL takes 18 minutes and 46 seconds to generate CGs 
for ten dependency sets, 
whereas the proposed approach takes 13 minutes and 8 seconds. 
There are 204 common dependencies 
in the dependency sets that we selected.
These common dependencies enable us 
to remove 203 redundant CG construction and make CG pool construction faster.
Another use case scenario that we present in the Table~\ref{tab:deps} is on-demand analysis.
The last row of the table shows that for serving on-demand CG generation, 
we decrease the time to 8 minutes and 34 seconds.
That happens only when there is no new dependency to be inserted 
in the CG pool in the request. 
Hence, we can fetch all dependencies from CG Pool without generating any new CG for dependencies.

We manually compared the CGs of OPAL and Stitching on a set of test cases that 
cover all Java language features~\cite{reif2019judge}. 
The results show that 
the soundness and precision of stitched CGs are the same as CGs solely generated by OPAL. 
We also plan to do the same manual analysis on a small subset of Maven packages in the future.

It is also worth mentioning that reported improvements 
are calculated on a small sample set and will be more tangible on a larger one. 
Hence, we plan to evaluate the approach on larger samples in the future.

\bibliographystyle{IEEEtran}
\bibliography{paper}

\end{document}